\documentclass{article}
\usepackage{amssymb}
\usepackage{amsmath}

\input{tcilatex}

\begin{document}

\title{ The flux of noncommutative $U(1)$ instanton through the fuzzy spheres%
}
\author{A.A.Henni\thanks{%
h\_moubine@yahoo.fr} and M.Lagraa\thanks{%
m.lagraa@mailcity.lycos.com} \\
Laboratoire de physique theorique\\
Universit\'{e} d'Oran Es-Senia, 31100, Algerie.}
\maketitle

\begin{abstract}
From the ADHM construction on noncommutative $\mathbb{R}_{\theta }^{4}$
vector space we investigate different $U(1)$ instanton solutions tied by
partial\ isometry trasformations. We recast these solutions under a form of
vector fields in noncommutative $\mathbb{R}_{\theta }^{3}$ vector space
which makes possible the calculus of their fluxes through fuzzy spheres. For
this end, we establish the noncommutative analog of Gauss theorem from which
we show that the flux of the $U(1)$ instantons through fuzzy spheres does
not depend on the radius of these spheres and it is invariant under partial
isometry transformations.

\textbf{PACS\ NUMBER:} 11.10.Nx, 11.15.Tk.
\end{abstract}

\textbf{Introduction}

Noncommutative geometry is a generalization of the usual differential
geometry in the sense that the usual description of manifolds by their
corresponding algebra\ of functions is reformulated by using noncommutive
algebras which are considered as algebras on noncommutative spaces \cite%
{connes,Madore,Gracia}. In physics one can hope that noncommutative geometry
gives alternatives to solve many problems such as renormalization of quantum
field theories where the fuzzy spheres are used in the regularization scheme 
\cite{Grosse}, quantization of gravity \cite{Doplicher}, superstring and
M-theory \cite{Banks, Douglas, Mtheory2} and quantum Hall effect \cite%
{Bellissard,Ezawa}.

In the several last years a great variety of works in field theories on
noncommutative geometry have been developed. In particular the Yang-Mills
gauge theories which are emerged from certain low energy limit of string
theory \cite{Schomerus,Seiberg} or from M theory compactification \cite%
{Douglas, Mtheory2}. In the most part of this development are treated some
non perturbative aspects of noncommutative gauge theories especially to
describe noncommutative instantons and their topological charge \cite%
{Nekrasov0,furuuchi1,furuuchi2,furuuchi3,furuuchi4,Masachi,Nekrasov2,Sako}\ .

In this work we mainly treat the invariance of the one topological charge
under partial isometry transformations of $U(1)$ instantons on
noncommutative $\mathbb{R}_{\theta }^{4}$ and establish the analog of Gauss
theorem in noncommutative $\mathbb{R}_{\theta }^{3}$ space from which we
show that the flux of $U(1)$ instantons through fuzzy spheres does not
depend on their radius (does not depend on the Hilbert space representations
of different fuzzy sphere algebras).

We begin this paper by recalling in section 1 some properties of
noncommutative $\mathbb{R}_{\theta }^{4}$ space and review briefly in
section 2\ the\ \textbf{ADHM} (Atiyah-Drinfeld-Hitchin-Manin) construction \
of instantons\ on this noncommutative space \cite{ADHM, Corrigan, Rajaraman,
Nekrasov0}.

In the third section, explicit solutions deduced from partial isometries
are\ investigated. We show that these partial isometries acts as
noncommutative $U(1)$ gauge transformations leaving invariant the instanton
number.

In section 4 we recast the $U(1)$-one-instantons in terms of operator
algebra over $\mathbb{R}_{\theta }^{3}$. This will leads us to view the
strength field of the $U(1)$ instanton as a vector field in $\mathbb{R}%
_{\theta }^{3}$. Finally we establish in the last section a noncommutative
analog to the Gauss theorem to calculate the flux of the $U(1)$ instanton
fields through fuzzy spheres then we show its invariance under partial
isomerty\ transformations.

\section{Noncommutative $\mathbb{R}_{\protect\theta }^{4}$}

In this section we consider $\mathbb{R}_{\theta }^{4}$ noncommutative space
which can be described by the complex coordinates

\begin{equation*}
\widehat{z}_{1}=\widehat{x}_{2}+i\widehat{x}_{1},\text{ \ \ \ \ }\widehat{z}%
_{2}=\widehat{x}_{4}+i\widehat{x}_{3},
\end{equation*}%
satisfying the commutation relations : 
\begin{equation*}
\left[ \widehat{z}_{\alpha },\widehat{\overline{z}}_{\beta }\right]
=-2\delta _{\alpha ,\beta }\theta ,\text{ }\left[ \widehat{z}_{\alpha },%
\widehat{z}_{\beta }\right] =0\text{ \ },(\alpha ,\beta =1,2)
\end{equation*}

\ \ The Hilbert space representation $\mathcal{H}$, on which the operators $%
\widehat{z}_{\alpha },\widehat{\overline{z}}_{\alpha }$ act, is spanned by
the basis

\begin{equation*}
\ \ \left\vert n_{1},n_{2}\right\rangle =\frac{(\widehat{z}_{1}/\sqrt{%
2\theta })^{n_{1}}}{\sqrt{n_{1}!}}\frac{(\widehat{z}_{2}/\sqrt{2\theta }%
)^{n_{2}}}{\sqrt{n_{2}!}}\left\vert 0,0\right\rangle ,\ \widehat{\overline{z}%
}_{\alpha }\left\vert 0,0\right\rangle =0
\end{equation*}

where the states $\left\vert n_{1},n_{2}\right\rangle $ are orthonormalized 
\begin{equation}
\langle n_{1},n_{2}\left\vert m_{1},m_{2}\right\rangle =\delta
_{n_{1}m_{1}}\delta _{n_{2}m2}.
\end{equation}

$\widehat{z}_{\alpha }$ and $\widehat{\overline{z}}_{\alpha }$ act on theses
states as:

\begin{eqnarray*}
\widehat{z}_{1}\left\vert n_{1},n_{2}\right\rangle &=&\sqrt{2\theta \left(
n_{1}+1\right) }\left\vert n_{1}+1,n_{2}\right\rangle ,\text{ \ \ \ }%
\widehat{z}_{2}\left\vert n_{1},n_{2}\right\rangle =\sqrt{2\theta \left(
n_{2}+1\right) }\left\vert n_{1},n_{2}+1\right\rangle , \\
\widehat{\overline{z}}_{1}\left\vert n_{1},n_{2}\right\rangle &=&\sqrt{%
2\theta n_{1}}\left\vert n_{1}-1,n_{2}\right\rangle ,\text{ \ \ \ }\widehat{%
\overline{z}}_{2}\left\vert n_{1},n_{2}\right\rangle =\sqrt{2\theta n_{2}}%
\left\vert n_{1},n_{2}-1\right\rangle .
\end{eqnarray*}

The operator algebra over $\mathbb{R}_{\theta \text{ }}^{4}$, denoted by $%
\widehat{\mathcal{A}}_{\theta }^{4}$, can also be described by the algebra $%
\mathcal{A}_{\theta }^{4}$ of c-number functions $f(z,\overline{z})$ endowed
with the normal ordering star product \cite{Alexianian} 
\begin{equation}
f(z,\overline{z})\ast g(z,\overline{z})=e^{2\theta \frac{\partial }{\partial 
\overline{z}}\frac{\partial }{\partial z^{\prime }}}f(z,\overline{z}%
)g(z^{\prime },\overline{z}^{\prime })_{\mid z^{\prime }=z,\overline{z}%
^{\prime }=\overline{z}}.  \label{star4}
\end{equation}%
The correspondence between the operators and the c-number functions is given
by:%
\begin{equation}
:\widehat{f}(\widehat{z}_{\alpha ,}\widehat{\overline{z}}_{\alpha
}):\longleftrightarrow f(z_{\alpha },\overline{z}_{\alpha })=\langle 
\overline{z}\mid :\widehat{f}(\widehat{z}_{\alpha ,}\widehat{\overline{z}}%
_{\alpha }):\mid \overline{z}\rangle \text{ ,}  \label{normalcorrespond}
\end{equation}%
where%
\begin{equation*}
\left\vert \overline{z}\right\rangle =\exp \left( -\frac{z_{\alpha }\widehat{%
\overline{z}}_{\alpha }}{4\theta }\right) \exp \left( \frac{\overline{z}%
_{\alpha }\widehat{z}_{\alpha }}{2\theta }\right) \left\vert 0,0\right\rangle
\end{equation*}%
is the coherent state satisfying $\widehat{\overline{z}}_{\alpha }\left\vert 
\overline{z}\right\rangle =\overline{z}_{\alpha }\left\vert \overline{z}%
\right\rangle $ and $\langle \overline{z}\left\vert \overline{z}%
\right\rangle =1$.

The derivatives $\partial _{\alpha }=\frac{\partial }{\partial z_{\alpha }}$
and $\partial _{\overline{\alpha }}=\frac{\partial }{\partial \overline{z}%
_{\alpha }}$ satisfy Leibnitz rules with respect to this star product (\ref%
{star4}). In the operator description, these derivatives are defined as 
\begin{equation}
\widehat{\partial }_{\alpha }\left( \cdot \right) =\frac{1}{2\theta }\left[ 
\widehat{\overline{z}}_{\alpha },\cdot \right] \text{, \ \ \ \ \ \ }\widehat{%
\partial }_{\overline{\alpha }}\left( \cdot \right) =\frac{-1}{2\theta }%
\left[ \widehat{z}_{\alpha },\cdot \right] .
\end{equation}

For the integration, the correspondence between the operator description and
the c-number function is given by

\begin{equation}
\int d^{4}xf\longleftrightarrow (2\pi \theta )^{2}Tr_{\mathcal{H}}(\widehat{f%
})  \label{integ defi}
\end{equation}%
where the trace of the operator is over the Hilbert space representation $%
\mathcal{H}$.

\section{Instantons and\ \textbf{ADHM} construction}

Instantons are localized finite-action non-perturbative solutions for the
Euclidian equations of motion of Yang-Mills gauge theories. In this section
we will recall the basic algoritm of \textbf{ADHM} construction \cite%
{Nekrasov0} to get instanton solutions in noncommutative space$\mathbb{R}%
_{\theta }^{4}$ which are just deformed versions of the commutative one \cite%
{ADHM, Corrigan, Rajaraman}.

The different steps of the ADHM construction for $U(N)$ $k$-instantons can
be summarized as follows

\begin{enumerate}
\item Solve the deformed ADHM equations 
\begin{equation*}
\begin{array}{l}
\text{ \ \ }\left[ B_{1},B_{1}^{\dagger }\right] +\left[ B_{2},B_{2}^{%
\dagger }\right] +II^{\dagger }-J^{\dagger }J=4\theta id_{k}. \\ 
\text{ \ \ }\left[ B_{1},B_{2}\right] +IJ=0.%
\end{array}%
\end{equation*}%
where $B_{1},B_{2}\in End\left( C^{k}\right) ,I\in Hom\left(
C^{N},C^{k}\right) $ and $J\in $ $Hom\left( C^{k},C^{N}\right) $

\item Define Dirac operator $\mathcal{D}_{z}$:$\left( C^{k}\oplus
C^{k}\oplus C^{N}\right) \otimes \widehat{\mathcal{A}}_{\theta
}^{4}\longrightarrow $ $\left( C^{k}\oplus C^{k}\right) \otimes \widehat{%
\mathcal{A}}_{\theta }^{4}$%
\begin{equation}
\mathcal{D}_{z}=\left( 
\begin{array}{ccc}
B_{2}-\widehat{z}_{2} & B_{1}-\widehat{z}_{1} & I \\ 
-(B_{1}^{\dagger }-\widehat{\overline{z}}_{1}) & B_{2}^{\dagger }-\widehat{%
\overline{z}}_{2} & J^{\dagger }%
\end{array}%
\right) .  \label{DDD}
\end{equation}

\item Look for all the $N$ orthonormalized\ solutions $\Psi ^{a}:\widehat{%
\mathcal{A}}_{\theta }^{4}\longrightarrow \left( C^{k}\oplus C^{k}\oplus
C^{N}\right) \otimes \widehat{\mathcal{A}}_{\theta }^{4}$ (the zero-modes)
to the equation 
\begin{equation*}
\mathcal{D}_{z}\Psi ^{a}=0,\text{\ \ }\Psi ^{a\dagger }\Psi ^{b}=\delta
^{ab}.
\end{equation*}

\item Construct the $U(N)$ gauge field 
\begin{equation*}
\widehat{A}=\Psi ^{\dagger }d\Psi .
\end{equation*}

The field strength of the gauge field $\widehat{A}$ is given by $\widehat{F}%
=d\widehat{A}+\widehat{A}^{2}$ and the topological number is defined by 
\begin{equation}
\mathcal{K}=\frac{-1}{16\pi ^{2}}\left( 2\pi \theta \right) ^{2}Tr\left( 
\widehat{F}\right) ^{2}  \label{number}
\end{equation}%
where the trace is taken both on the group indices (for the general case of
the $U(N)$ gauge group) and on the Hilbert space $\mathcal{H}.$ The formula (%
\ref{number}) is the noncommutative version of the second Chern character
defined by: 
\begin{equation*}
\mathcal{K}=\frac{-1}{16\pi ^{2}}\int d^{4}xTr_{U(N)}\left( F_{\mu \nu }%
\widetilde{F}^{\mu \nu }\right)
\end{equation*}%
where $\widetilde{F}^{\mu \nu }=\ast F^{\mu \nu }=\frac{1}{2}\epsilon ^{\mu
\nu \rho \sigma }F^{\rho \sigma }$ is the dual of the strength field.
\end{enumerate}

\section{$U(1)$-one-Instanton and partial isometries}

We will now\ concentrate on the partial isometry transformations of the
known example of $U(1)$-one-instanton solution on $\mathbb{R}_{\theta }^{4}$%
(see\cite{furuuchi1}). \ For the $U(1)$-one-instanton solutions on $\mathbb{R%
}_{\theta }^{4}$ , $k=N=1$, thus the matrices $B_{1,2},I,J$ become complex
numbers. Inserting this in the ADHM equations we obtain $II^{\dagger
}-J^{\dagger }J=4\theta $ and $IJ=0$, we take $I=2\sqrt{\theta }$,\ $J=0.$%
The parameters $B_{1,2}$ are interpreted as the position of the instanton.
Due to the translational invariance of $\mathbb{R}_{\theta }^{4}$ we can
consider the case where the instanton is localized at the origin i.e. $%
B_{1,2}=0$. In this case the Dirac operator reads 
\begin{equation}
\mathcal{D}_{z}=\left( 
\begin{array}{c}
\begin{array}{ccc}
-\widehat{z}_{2} & -\widehat{z}_{1} & 2\sqrt{\theta }%
\end{array}
\\ 
\begin{array}{ccc}
\widehat{\overline{z}}_{1} & -\widehat{\overline{z}}_{2} & 0%
\end{array}%
\end{array}%
\right) .  \label{Dirac}
\end{equation}%
Finely we look for the solution $\Psi $ of the Dirac equation\ 
\begin{equation}
\mathcal{D}_{z}\Psi =0  \label{Dequ}
\end{equation}%
whose solution is given by : 
\begin{equation}
\psi _{0}=\left( 
\begin{array}{c}
2\sqrt{\theta }\widehat{\overline{z}}_{2} \\ 
2\sqrt{\theta }\widehat{\overline{z}}_{1} \\ 
\widehat{z}\widehat{\overline{z}}%
\end{array}%
\right) \frac{1}{\sqrt{\widehat{z}\widehat{\overline{z}}(\widehat{z}\widehat{%
\overline{z}}+4\theta )}}  \label{simple}
\end{equation}%
\ where $\widehat{z}\widehat{\overline{z}}=\widehat{z}_{1}\widehat{\overline{%
z}}_{1}+\widehat{z}_{2}\widehat{\overline{z}}_{2}$. We can see that this
solution is not normalized as said in the ADHM data, because of the operator 
$\frac{1}{\sqrt{\widehat{z}\widehat{\overline{z}}(\widehat{z}\widehat{%
\overline{z}}+4\theta )}}$\ which indefinite on the vacuum\ state $%
\left\vert 0,0\right\rangle $. However this solution is well normalized in
the subspace where the state $\left\vert 0,0\right\rangle $ is projected
out. The subspace $\mathcal{H-}\left\{ \left\vert 0,0\right\rangle \right\} $%
\ will be denoted by $\mathcal{H}_{00}$.

The gauge field is directly calculated\ in $\mathcal{H}_{00}$ by%
\begin{equation}
\widehat{A}_{00}=\psi _{0}^{\dagger }\frac{1}{2\theta }\left[ \widehat{%
\overline{z}}^{i},\psi _{0}\right] dz^{i}-\psi _{0}^{\dagger }\frac{1}{%
2\theta }\left[ \widehat{z}^{i},\psi _{0}\right] d\overline{z}^{i}.
\label{AZERO}
\end{equation}

By using the relations $\widehat{z}_{\alpha }\widehat{f}\left( \widehat{z}%
\widehat{\overline{z}}\right) =\widehat{f}\left( \widehat{z}\widehat{%
\overline{z}}-2\theta \right) \widehat{z}_{\alpha },$ $\widehat{\overline{z}}%
_{\alpha }\widehat{f}\left( \widehat{z}\widehat{\overline{z}}\right) =%
\widehat{f}\left( \widehat{z}\widehat{\overline{z}}+2\theta \right) \widehat{%
\overline{z}}_{\alpha }$ and the solution (\ref{simple}) we may compute
explicitly (\ref{AZERO}) to get:

\begin{equation}
\widehat{A}_{00}=\frac{1}{2\theta }((\frac{\widehat{z}\widehat{\overline{z}}(%
\widehat{z}\widehat{\overline{z}}+6\theta )}{(\widehat{z}\widehat{\overline{z%
}}+2\theta )(\widehat{z}\widehat{\overline{z}}+4\theta )})^{\frac{1}{2}}-1)%
\widehat{\overline{z}}_{\alpha }dz^{\alpha }-h.c.  \label{AZEROZ}
\end{equation}%
leading to a strength field $\widehat{F}_{00}=d\widehat{A}_{00}+\widehat{A}%
_{00}\cdot \widehat{A}_{00}$ of the form

\begin{eqnarray}
\widehat{F}_{00} &=&-\frac{8\theta }{\widehat{z}\widehat{\overline{z}}(%
\widehat{z}\widehat{\overline{z}}+2\theta )(\widehat{z}\widehat{\overline{z}}%
+4\theta )}(\widehat{\overline{z}}_{1}\widehat{z}_{2})dz_{1}d\overline{z}%
_{2}-(\widehat{z}_{1}\widehat{\overline{z}}_{2})\frac{8\theta }{\widehat{z}%
\widehat{\overline{z}}(\widehat{z}\widehat{\overline{z}}+2\theta )(\widehat{z%
}\widehat{\overline{z}}+4\theta )}dz_{2}d\overline{z}_{1}  \notag \\
&&-\frac{4\theta (\widehat{z}_{1}\widehat{\overline{z}}_{1}-\widehat{z}_{2}%
\widehat{\overline{z}}_{2})}{\widehat{z}\widehat{\overline{z}}(\widehat{z}%
\widehat{\overline{z}}+2\theta )(\widehat{z}\widehat{\overline{z}}+4\theta )}%
dz_{1}d\overline{z}_{1}+\frac{4\theta (\widehat{z}_{1}\widehat{\overline{z}}%
_{1}-\widehat{z}_{2}\widehat{\overline{z}}_{2})}{\widehat{z}\widehat{%
\overline{z}}(\widehat{z}\widehat{\overline{z}}+2\theta )(\widehat{z}%
\widehat{\overline{z}}+4\theta )}dz_{2}d\overline{z}_{2}  \label{S3}
\end{eqnarray}

which exhibits the anti-self duality conditions : 
\begin{equation*}
\widehat{F}_{00z_{1}\overline{z}_{1}}=-\widehat{F}_{00z_{2}\overline{z}_{2}},%
\text{ \ \ \ }\widehat{F}_{00z_{1}z_{2}}=0.
\end{equation*}%
The topological charge of the $U(1)$ instanton expressed like in relation (%
\ref{number}) by using 
\begin{eqnarray}
\left( F_{00}\right) ^{2} &=&\left[ 16\left( \widehat{F}_{00z_{1}\overline{z}%
_{1}}\widehat{F}_{00z_{2}\overline{z}_{2}}-\frac{1}{2}\left( \widehat{F}%
_{00z_{1}\overline{z}_{2}}\widehat{F}_{00z_{2}\overline{z}_{1}}+\widehat{F}%
_{00z_{2}\overline{z}_{1}}\widehat{F}_{00z_{1}\overline{z}_{2}}\right)
\right) \right]  \notag \\
&=&-\frac{\left( 16\theta \right) ^{2}}{\widehat{z}\widehat{\overline{z}}(%
\widehat{z}\widehat{\overline{z}}+2\theta )(\widehat{z}\widehat{\overline{z}}%
+4\theta )}  \label{F002}
\end{eqnarray}%
to get \cite{furuuchi1}%
\begin{eqnarray*}
\mathcal{K} &=&\frac{\theta ^{2}}{4}Tr_{\mathcal{H}_{00}}\left( \frac{\left(
16\theta \right) ^{2}}{\widehat{z}\widehat{\overline{z}}(\widehat{z}\widehat{%
\overline{z}}+2\theta )(\widehat{z}\widehat{\overline{z}}+4\theta )}\right) =%
\underset{\overrightarrow{n}\neq \overrightarrow{0}}{\sum }\frac{4}{%
(n_{1}+n_{2})(n_{1}+n_{2}+1)^{2}(n_{1}+n_{2}+2)} \\
&=&4\underset{n\neq 0}{\sum }\frac{1}{n(n+1)(n+2)}=1.
\end{eqnarray*}%
where we denote the ordered couple $\left( n_{1},n_{2}\right) $ by $%
\overrightarrow{n}$\ and the sum $n_{1}+n_{2}$ by $n$.

The investigation of the transformed solution by partial isomery is deduced
from the fact that the solution

\begin{equation*}
\widetilde{\psi }_{0}=\left( 
\begin{array}{c}
2\sqrt{\theta }\widehat{\overline{z}}_{2} \\ 
2\sqrt{\theta }\widehat{\overline{z}}_{1} \\ 
\widehat{z}\widehat{\overline{z}}%
\end{array}%
\right)
\end{equation*}%
of (\ref{Dequ}) belong to a right module. Then we may also consider others
solutions of the form $\psi =\widetilde{\psi }_{0}(\widehat{\overline{z}}%
_{1})^{K}(\widehat{\overline{z}}_{2})^{L}$ $(K\geq 0,$ $L\geq 0)$. The
normalization of these solutions gives

\begin{equation}
\psi =\psi _{0}U=\left( 
\begin{array}{c}
2\sqrt{\theta }\widehat{\overline{z}}_{2} \\ 
2\sqrt{\theta }\widehat{\overline{z}}_{1} \\ 
\widehat{z}\widehat{\overline{z}}%
\end{array}%
\right) \frac{1}{\sqrt{\widehat{z}\widehat{\overline{z}}(\widehat{z}\widehat{%
\overline{z}}+4\theta )}}\left( \widehat{\overline{z}}_{1}\frac{1}{\sqrt{%
\widehat{z}_{1}\widehat{\overline{z}}_{1}}}\right) ^{K}\left( \widehat{%
\overline{z}}_{2}\frac{1}{\sqrt{\widehat{z}_{2}\widehat{\overline{z}}_{2}}}%
\right) ^{L}  \label{NORMSOLU}
\end{equation}%
where $U$ is given by

\begin{equation}
U=\left( \widehat{\overline{z}}_{1}\frac{1}{\sqrt{\widehat{z}_{1}\widehat{%
\overline{z}}_{1}}}\right) ^{K}\left( \widehat{\overline{z}}_{2}\frac{1}{%
\sqrt{\widehat{z}_{2}\widehat{\overline{z}}_{2}}}\right) ^{L}.
\label{transformS}
\end{equation}

One can check that (\ref{NORMSOLU}) can be rewritten as 
\begin{equation}
\psi =\widetilde{\psi }_{0}\widehat{\overline{z}}_{1}^{K}\widehat{\overline{z%
}}_{2}^{L}N(\widehat{z},\widehat{\overline{z}})^{-\frac{1}{2}}
\end{equation}%
where

\begin{equation*}
N(\widehat{z},\widehat{\overline{z}})=(\widehat{z}\widehat{\overline{z}}%
-2\theta (K+L))(\widehat{z}\widehat{\overline{z}}-2\theta (K+L-2))\underset{%
k=1}{\overset{K}{\Pi }}(\widehat{z}_{1}\widehat{\overline{z}}_{1}-2\theta
(k-1))\overset{L}{\underset{l=1}{\Pi }}(\widehat{z}_{2}\widehat{\overline{z}}%
_{2}-2\theta (l-1)).
\end{equation*}

We will restrict ourselves, for the same reason explained above, on the
Hilbert subspace states $\mathcal{H}_{KL}$ on which the solution is well
defined and well normalized i.e., the eigenstates of $N(\widehat{z},\widehat{%
\overline{z}})$ whose the eigenvalues do not vanish. These nonsingular
states are $\left\vert n_{1,}n_{2}\right\rangle $ such that\ $n_{1}\geqslant
K,n_{2}\geqslant L$, and $(n_{1},n_{2})\neq (K,L)$. This basis span the
subspace $\mathcal{H}_{KL}=P_{KL}\mathcal{H}$ where the projector\ $P_{KL}$
reads%
\begin{eqnarray*}
P_{KL} &=&1-\overset{K}{\underset{n_{1}=0}{\sum }}\underset{n_{2}=0}{\overset%
{L}{\sum }}\left\vert n_{1},n_{2}\right\rangle \left\langle
n_{1},n_{2}\right\vert -\overset{K}{\underset{n_{1}=0}{\sum }}\underset{%
n_{2}=L+1}{\overset{\infty }{\sum }}\left\vert n_{1},n_{2}\right\rangle
\left\langle n_{1},n_{2}\right\vert \\
&&-\overset{\infty }{\underset{n_{1}=K+1}{\sum }}\underset{n_{2}=0}{\overset{%
L}{\sum }}\left\vert n_{1},n_{2}\right\rangle \left\langle
n_{1},n_{2}\right\vert .
\end{eqnarray*}

Before we study this solution, we investigate some properties of $U$
transformations which we will see below that they act like noncommutative $%
U(1)$\ gauge transformations. They are not unitary but satisfy the
properties of partial isometries\footnote{%
The transformation $U$\ correspond to a multiplication by a phase factor in
the classical limit ($\theta \longrightarrow 0$). In this limit $z_{i},%
\overline{z}_{i}$ become usual complexe coordinates, then if we put $%
z_{i}=r_{i}e^{i\varphi _{i}}$ (polar coordinates $r_{i}$ ,$\varphi _{i}$\ )
it is obvious to have $\left( \overline{z}_{i}\frac{1}{\sqrt{z_{i}\overline{z%
}_{i}}}\right) ^{J}=\left( e^{i\varphi _{i}}\right) ^{J}$, thus the
transformation take the usual form $U=e^{i\Phi }$, where $\Phi =K\varphi
_{1}+L\varphi _{2}$.}. 
\begin{eqnarray}
UU^{\dagger } &=&id_{\mathcal{H}}\text{, \ \ \ \ \ \ }U^{\dagger }U=P_{KL},
\label{Urelation} \\
P_{KL}U^{\dagger } &=&U^{\dagger }\text{, \ \ \ \ \ \ \ }UP_{KL}=U.  \notag
\end{eqnarray}

In $\mathcal{H}_{KL}$, the relations (\ref{Urelation}) read 
\begin{equation}
UU^{\dagger }=U^{\dagger }U=id_{\mathcal{H}_{KL}}.  \label{usual}
\end{equation}

Under $U$, the gauge field transforms in this subspace as 
\begin{eqnarray}
\widehat{A}_{KL} &=&\psi ^{\dagger }d\psi =U^{\dagger }\psi _{0}^{\dagger
}d(\psi _{0})U+U^{\dagger }\psi _{0}^{\dagger }\psi _{0}dU  \notag \\
&=&U^{\dagger }\widehat{A}_{00}U+U^{\dagger }dU  \label{gaugeF}
\end{eqnarray}%
where we have used the fact that in $\mathcal{H}_{KL}$ we have $U^{\dagger
}\psi _{0}^{\dagger }\psi _{0}=U^{\dagger }$. We can see from the relation (%
\ref{gaugeF}) that $U$ acts (from\ $\mathcal{H}_{00}$ to $\mathcal{H}_{KL}$)
like noncommutative $U(1)$-gauge transformations act on the gauge fields.

Now the strength field in $\mathcal{H}_{KL}$ is defined as usual by: 
\begin{equation}
\widehat{F}_{KL}=d\widehat{A}_{KL}+\widehat{A}_{KL}\widehat{A}_{KL}.
\label{UFKL}
\end{equation}%
Using (\ref{gaugeF}), $dU^{\dagger }U=-U^{\dagger }dU$ and\ \ $dUU^{\dagger
}=-UdU^{\dagger }$ in $\mathcal{H}_{KL}$, we get%
\begin{equation}
\widehat{F}_{KL}=U^{\dagger }\widehat{F}_{00}U  \label{FKL}
\end{equation}%
which is the gauge transform of the strength field $\widehat{F}_{00}$ (in $%
\mathcal{H}_{00}$) in the subspace $\mathcal{H}_{KL}$. The instanton number
must be unchanged by the $U$-transformations, this is what we will see in
the calculus of 
\begin{equation}
\mathcal{K}=\frac{-1}{16\pi ^{2}}\left( 2\pi \theta \right) ^{2}Tr_{\mathcal{%
H}_{KL}}\left( \widehat{F}_{KL}\right) ^{2}=\frac{-1}{16\pi ^{2}}\left( 2\pi
\theta \right) ^{2}Tr_{\mathcal{H}_{KL}}\left( U^{\dagger }\left( \widehat{F}%
_{00}\right) ^{2}U\right) .  \label{last}
\end{equation}%
\ 

\bigskip In fact, by using (\ref{F002}) we get%
\begin{eqnarray}
\left( \widehat{F}_{KL}\right) ^{2} &=&U^{\dagger }\left( \widehat{F}%
_{00}\right) ^{2}U=-U^{\dagger }\frac{\left( 16\theta \right) ^{2}}{\widehat{%
z}\widehat{\overline{z}}(\widehat{z}\widehat{\overline{z}}+2\theta )^{2}(%
\widehat{z}\widehat{\overline{z}}+4\theta )}U  \notag \\
&=&-\frac{\left( 16\theta \right) ^{2}}{(\widehat{z}\widehat{\overline{z}}%
-2\theta \left( K+L\right) )(\widehat{z}\widehat{\overline{z}}-2\theta
\left( K+L-1\right) )^{2}(\widehat{z}\widehat{\overline{z}}-2\theta \left(
K+L-2\right) )}P_{KL}  \label{FKL2}
\end{eqnarray}%
Inserting (\ref{FKL2}) in (\ref{last}), we obtain%
\begin{eqnarray*}
\mathcal{K} &=&4\underset{n_{2}\geq L}{\underset{n_{1}\geq K}{\underset{%
\overrightarrow{n}\neq (K,L)}{\sum }}}\frac{1}{%
(n-(K+L))(n-(K+L-1))^{2}(n-(K+L-2))} \\
&=&4\underset{\overrightarrow{n}\neq (0,0)}{\sum }\frac{1}{n(n+1)^{2}(n+2)}%
=1.
\end{eqnarray*}%
which shows the invariance of the instanton number partial isometries.

\section{Reduction to the vector space $\mathbb{R}_{\protect\theta }^{3}$}

In this section we will calculate the flux of the $U(1)$-one instanton
strength field\ through the fuzzy sphere. This investigation is justified by
the fact that the $U(1)$ instanton components (\ref{S3}) can be written in
terms of the $\mathbb{R}_{\theta }^{3}$\ vector space coordinates $\widehat{x%
}^{i}$ as a three vector $\overrightarrow{F}$ whose components are defined
in $\mathcal{H}_{00}$ as:

\begin{equation*}
\widehat{F}_{00z_{1}\overline{z}_{1}}=\widehat{F}^{3}=\frac{-\theta \widehat{%
x}^{3}}{\widehat{x}^{0}(\widehat{x}^{0}+\theta )(\widehat{x}^{0}+2\theta )},
\end{equation*}

\begin{equation*}
\widehat{F}_{00z_{1}\overline{z}_{2}}=\widehat{F}^{+}=\frac{-\theta }{%
\widehat{x}^{0}(\widehat{x}^{0}+\theta )(\widehat{x}^{0}+2\theta )}\left( 
\widehat{x}^{1}+i\widehat{x}^{2}\right) =F^{1}+iF^{2},
\end{equation*}

\begin{equation}
\widehat{F}_{00z_{2}\overline{z}_{1}}=\widehat{F}^{-}=\frac{-\theta }{%
\widehat{x}^{0}(\widehat{x}^{0}+\theta )(\widehat{x}^{0}+2\theta )}\left( 
\widehat{x}^{1}-i\widehat{x}^{2}\right) =F^{1}-iF^{2}  \label{FR3}
\end{equation}%
with $\widehat{x}^{0}=\frac{1}{2}\widehat{z}\widehat{\overline{z}}$. The
coordinates on $\mathbb{R}_{\theta }^{3}$ are defined from those on $\mathbb{%
R}_{\theta }^{4}$ by\ (Hopf fibration) 
\begin{equation}
\widehat{x}^{i}=\frac{1}{2}\widehat{z}_{\alpha }\tau _{\alpha \beta }^{i}%
\widehat{\overline{z}}_{\beta },\text{ \ \ }i=1,2,3;\text{ \ \ }\alpha
,\beta =1,2  \label{reduction}
\end{equation}%
where $\tau ^{i}$ are the three Pauli matrices. The coordinates $\widehat{x}%
^{i}$ generate a subalgebra $\widehat{\mathcal{A}}_{\theta }^{3}\subset 
\widehat{\mathcal{A}}_{\theta }^{4}$ satisfying the commutation rules%
\begin{equation}
\left[ \widehat{x}^{i},\widehat{x}^{j}\right] =i2\theta \epsilon ^{ijk}%
\widehat{x}^{k},\ \ \left[ \widehat{x}^{0},\widehat{x}^{i}\right] =0
\label{FSPHERE}
\end{equation}%
and the relation%
\begin{equation*}
\widehat{x}^{i}\cdot \widehat{x}^{i}=\widehat{x}^{0}\left( \widehat{x}%
^{0}+2\theta \right)
\end{equation*}%
where the repeated indices are summed over. It is convenient to use, for the 
$\widehat{\mathcal{A}}_{\theta }^{3}$ algebra the Schwinger basis of the
Hilbert space 
\begin{equation*}
\left\vert J,m\right\rangle =\frac{\left( \frac{\widehat{z}_{1}}{\sqrt{%
2\theta }}\right) ^{J+m}}{\sqrt{\left( J+m\right) !}}\frac{\left( \frac{%
\widehat{z}_{2}}{\sqrt{2\theta }}\right) ^{J-m}}{\sqrt{\left( J-m\right) !}}%
\left\vert 0,0\right\rangle
\end{equation*}%
where $J=0,\frac{1}{2},1,...\infty \,$and $m$\ runs by integer steps over
the range $-$ $J\leq m\leq J$.

The algebra $\widehat{\mathcal{A}}_{\theta \text{ }}^{3}$can also be
described by the algebra $\mathcal{A}_{\theta }^{3}$ of c-number function $%
f(x)$ of $x^{i}=\frac{1}{2}z_{\alpha }\tau _{\alpha \beta }^{i}\overline{z}%
_{\beta }$ and $x^{i}\cdot x^{i}=\left( x^{0}\right) ^{2}$ endowed with the
star product \cite{Lagraa}

\begin{equation}
f(x)\ast g(x)=e^{\theta (x^{0}\delta ^{ij}+i\epsilon ^{ijk}x^{k})\frac{%
\partial }{\partial x^{i}}\frac{\partial }{y^{j}}}f(x)g(y)_{\mid y^{i}=x^{i}}%
\text{ }.  \label{star3}
\end{equation}%
deduced from the star product of $\mathcal{A}_{\theta }^{3}$ by restriction.

In terms of elements of the algebra $\mathcal{A}_{\theta }^{3}$ we can
decompose the strength field\ components (\ref{FR3}) as

\begin{eqnarray}
F_{J}^{i} &=&-\underset{J=\frac{1}{2}}{\overset{\infty }{\sum }}\frac{\theta 
}{x^{0}\ast \left( x^{0}+\theta \right) \ast \left( 2x^{0}+2\theta \right) }%
\ast x^{i}\ast P_{J}\left( x^{0}\right)  \notag \\
&=&-\frac{1}{4\theta ^{2}}\underset{J=\frac{1}{2}}{\overset{\infty }{\sum }}%
\frac{1}{J\left( J+1\right) \left( 2J+1\right) }x^{i}\ast P_{J}\left(
x^{0}\right)  \notag \\
&=&-\frac{1}{4\theta ^{2}}\underset{J=\frac{1}{2}}{\overset{\infty }{\sum }}%
\frac{1}{J\left( J+1\right) \left( 2J+1\right) }x^{i}P_{J-\frac{1}{2}}=%
\underset{J=\frac{1}{2}}{\overset{\infty }{\sum }}F_{J}^{i}
\label{fuzzyfield}
\end{eqnarray}%
where we have used $x^{0}\ast P_{J}\left( x^{0}\right) =2\theta JP_{J}\left(
x^{0}\right) $ to define ($x^{0})^{-1}\ast P_{J}\left( x^{0}\right) =\frac{1%
}{2\theta J}P_{J}\left( x^{0}\right) $ for $J\neq 0$ and $x^{i}\ast
P_{J}\left( x^{0}\right) =x^{i}P_{J-\frac{1}{2}}\left( x^{0}\right) .$ $%
P_{J}\left( x^{0}\right) =\frac{\left( \frac{x^{0}}{\theta }\right) ^{2J}}{%
\left( 2J\right) !}e^{-\frac{x^{0}}{\theta }}$ is the projector on the Fuzzy
sphere of radius $J$ corresponding to the operator projector $\widehat{P}%
_{J}=\overset{J}{\underset{m=-J}{\sum }}\left\vert J,m\right\rangle
\left\langle J,m\right\vert $.

Each term of the sum (\ref{fuzzyfield}), $F_{J}^{i}=\frac{-1}{J\left(
J+1\right) \left( 2J+1\right) }x^{i}\ast P_{J}\left( x^{0}\right) $ $=\frac{%
-1}{J\left( J+1\right) \left( 2J+1\right) }x_{J}^{i}$ corresponds to the
strength field defined on the fuzzy sphere algebra of radius $J$,$\
S_{\theta ,J}^{2}=P_{J}\ast \mathcal{A}_{\theta }^{3}\ast P_{J}\subset 
\mathcal{A}_{\theta }^{3}\subset \mathcal{A}_{\theta }^{4}$ generated by $%
x_{J}^{i}=P_{J}\ast x^{i}\ast P_{J}=x^{i}\ast P_{J}$.

The corresponding operator subalgebra $\widehat{S}_{\theta ,J}^{2}$
generated by $\widehat{x}_{J}^{i}=\widehat{P}_{J}\widehat{x}^{i}\widehat{P}%
_{J}=\widehat{x}^{i}\widehat{P}_{J}$\ satisfy the commutation relation rules
on the fuzzy sphere%
\begin{equation*}
\left[ \widehat{x}_{J}^{i},\widehat{x}_{J}^{j}\right] =i2\theta \epsilon
^{ijk}\widehat{x}^{k},\ \ \left[ \widehat{x}_{J}^{0},\widehat{x}_{J}^{i}%
\right] =0.
\end{equation*}%
and%
\begin{equation*}
\widehat{x}_{J}^{i}\cdot \widehat{x}_{J}^{i}=\widehat{x}_{J}^{0}\left( 
\widehat{x}_{J}^{0}+2\theta \right) =4\theta ^{2}J(J+1)\widehat{P}_{J}
\end{equation*}

The subalgebra $\widehat{S}_{\theta ,J}^{2}$ is realized in the Hilbert
subspace $\mathcal{H}_{J}\subset \mathcal{H}$ spanned by the basis $%
\left\vert J,m\right\rangle $, $-$ $J\leq m\leq J$.

\section{Flux through the Fuzzy sphere}

The form ($\overrightarrow{F}\approx \frac{\overrightarrow{x}}{(x^{0})^{3}}$%
) of the strength field (\ref{FR3}) permits us to view it as the
noncommutative analog of the classical Dirac magnetic monopole field (or by
duality as the static coulombian electric field). So it is interesting to
calculate the flux of this noncommutative vector field through a fuzzy
sphere. To perform the\ calculus of the flux of the vector fields through
fuzzy spheres, we generalize the technics of the Gauss theorem to the
noncommutative space by using the integration measures on $\mathbb{R}%
_{\theta }^{3}$ and on\ the fuzzy sphere investigated in \cite{Lagraa}.

For this end we start with the derivatives with respect to $z_{\alpha }$ and 
$\overline{z}_{\alpha }$ of a scalar field $\varphi \left( x\right) $\ $\in 
\mathcal{A}_{\theta }^{3}$ giving components $\partial _{\alpha }\varphi
\left( x\right) $, $\partial _{\bar{\alpha}}\varphi \left( x\right) $ $\in 
\mathcal{A}_{\theta }^{4}$ of a quadri-vector given as 
\begin{equation}
\partial _{\alpha }\varphi \left( x\right) =\partial _{\bar{\alpha}}\left(
x^{i}\right) \partial _{i}\varphi \left( x\right) =\frac{1}{2}\tau _{\alpha
\beta }^{i}\overline{z}_{\beta }\partial _{i}\varphi \left( x\right) ,\text{
\ \ }\partial _{\bar{\alpha}}\varphi \left( x\right) =\partial _{\bar{\alpha}%
}\left( x^{i}\right) \partial _{i}\varphi \left( x\right) =\frac{1}{2}%
z_{\beta }\tau _{\beta \alpha }^{i}\partial _{i}\varphi \left( x\right)
\label{TRANSF}
\end{equation}%
where $\partial _{i}=\frac{\partial }{\partial x^{i}}$. These relations are
similar to the transformations of the\ covariant components of a\
quadri-vector under a coordinate transformations. In our case the
transformation $z_{\alpha },$ $\overline{z}_{\alpha }\longrightarrow x^{i}=%
\frac{1}{2}z_{\alpha }\tau _{\alpha \beta }^{i}\overline{z}_{\beta }$ is a\
surjection from $\mathbb{R}_{\theta }^{4}$\ to $\mathbb{R}_{\theta }^{3}$
which is not invertible. In what follows, we\ only consider\ quadri-vectors $%
V$ of components $V_{\alpha }$, $V_{\bar{\alpha}}$ $\in \mathcal{A}_{\theta
}^{4}$verifying the same transformations like in (\ref{TRANSF}) 
\begin{equation}
V_{\alpha }(z,\overline{z})=\frac{1}{2}\tau _{\alpha \beta }^{i}\overline{z}%
_{\beta }V_{i}(x),\text{ \ }V_{\bar{\alpha}}(z,\overline{z})=\frac{1}{2}%
z_{\beta }\tau _{\beta \alpha }^{i}V_{i}(x).  \label{vector transform}
\end{equation}%
where $V^{i}\left( x\right) \in \mathcal{A}_{\theta }^{3}$ are the
components of a three vector $\overrightarrow{V}\left( x\right) $ of $%
\mathbb{R}_{\theta }^{3}$.\ The quadri-divergence of (\ref{vector transform}%
) is given in terms of three-divergence of $\overrightarrow{V}\left(
x\right) $ as

\begin{equation}
\partial _{\bar{\alpha}}V_{\alpha }+\partial _{\alpha }V_{\bar{\alpha}%
}=x^{0}\partial _{i}V_{i}\in \mathcal{A}_{\theta }^{3}  \label{DIVV}
\end{equation}%
where we have used $Tr\left( \tau \right) =0$\ and the relation $\frac{1}{4}%
z_{\alpha }\tau _{\alpha \rho }^{i}\tau _{\rho \beta }^{j}\overline{z}%
_{\beta }=\frac{1}{2}(x^{0}\delta ^{ij}+i\epsilon ^{ijk}x^{k}).$The
integration of the above divergence over $\mathbb{R}_{\theta }^{4}$ reduces
to the integration over\ $\mathbb{R}_{\theta }^{3}$. In fact for a
coordinate system of the form%
\begin{equation*}
z_{1}=R\cos \frac{\theta }{2}e^{i\frac{\psi +\varphi }{2}}\text{ \ \ , \ \ \ 
}z_{2}=R\sin \frac{\theta }{2}e^{i\frac{\psi -\varphi }{2}}
\end{equation*}%
the measure on $\mathbb{R}_{\theta }^{4}$ reads%
\begin{equation*}
dz_{1}d\overline{z}_{1}dz_{2}d\overline{z}_{2}=\frac{R^{2}}{2}d\left( \frac{%
R^{2}}{2}\right) \sin \theta d\theta d\varphi d\psi =x^{0}dx^{0}d\Omega d\psi
\end{equation*}%
where $\frac{R^{2}}{2}=\frac{z\overline{z}}{2}=x^{0}.$ Since the elements of 
$\mathcal{A}_{\theta }^{3}\subset \mathcal{A}_{\theta }^{4}$ do not depend
on the angle $\psi $, their integration over $\mathbb{R}_{\theta }^{4}$
factorize into an integration over $\mathbb{R}_{\theta }^{4}$ as $\int
x^{0}dx^{0}d\Omega \mathcal{A}_{\theta }^{3}$ times an integration along the
angle $\psi $, $\int d\psi =\pi $. Then the measure over the $\mathbb{R}%
_{\theta }^{4}$ is given by $d^{3}x_{\theta }=x^{0}dx^{0}d\Omega $ which
differs from the usual $d^{3}x$ by factors of $\frac{1}{x^{0}}$. The
integration of (\ref{DIVV}) in $\mathbb{R}_{\theta }^{3}$ is given by 
\begin{equation}
\int \frac{d\overline{z}_{1}dz_{1}d\overline{z}_{2}dz_{2}}{\pi }\left(
\partial _{\bar{\alpha}}V_{\alpha }+\partial _{\alpha }V_{\bar{\alpha}%
}\right) =\int d^{3}x_{\theta }x^{0}\partial _{i}V_{i}=\int d^{3}x\partial
_{i}V_{i}.  \label{divintegral}
\end{equation}

This relation suggests that for the integration over $\mathbb{R}_{\theta
}^{3}$, the correspondence between the operator description and the c-number
functions is given by 
\begin{equation*}
\int \frac{d^{3}x}{x^{0}}f(x)\longleftrightarrow \frac{\left( 2\pi \theta
\right) ^{2}}{\pi }Tr_{\mathcal{H}}\left( \widehat{f}(\widehat{x})\right)
\end{equation*}%
where $f(x)\in \mathcal{A}_{\theta }^{3}$ and $\widehat{f}(\widehat{x})\in 
\widehat{\mathcal{A}}_{\theta }^{3}$. Now if we take $V^{i}=F^{i}$, the
divergence\footnote{%
Since the derivative $\partial _{i}$ is not a proper derivative with respect
to the star-prduct \cite{Lagraa}, we perform the calculus of the\
star-products then use the derivatives $\partial _{i}$ on the result which
is a c-number function of $x^{i}$.}\ $\partial _{i}F^{i}$ reads%
\begin{equation*}
\partial _{i}F^{i}=-\frac{1}{4\theta ^{2}}\overset{\infty }{\underset{J=%
\frac{1}{2}}{\sum }}\left( \frac{2P_{J-\frac{1}{2}}\left( x^{0}\right) }{%
(J)(2J+1)}-\frac{2P_{J}\left( x^{0}\right) }{(J+1)(2J+1)}\right)
\end{equation*}%
Its integration over a\ volume in $\mathbb{R}_{\theta }^{3}$ bounded by two
fuzzy spheres of radii $J_{2}>J_{1}$ respectively $(J_{1}\neq 0)$ reads%
\begin{equation*}
\Phi =\int d^{3}x\underset{J=J_{1}}{\overset{J_{2}}{\sum }}\partial
_{i}F_{J}^{i}=-\frac{1}{4\theta ^{2}}\int d^{3}x\overset{J_{2}}{\underset{%
J=J_{1}}{\sum }}\left( \frac{2P_{J-\frac{1}{2}}\left( x^{0}\right) }{%
(J)(2J+1)}-\frac{2P_{J}\left( x^{0}\right) }{(J+1)(2J+1)}\right)
\end{equation*}%
The contribution of the first term for $J+\frac{1}{2}$ automatically cancel
those of the second term of $J$ to leave the contributions at the boundaries
as%
\begin{equation}
\Phi =-\frac{4\pi }{4\theta ^{2}}\int \left( x^{0}\right) ^{2}dx^{0}\overset{%
J_{2}}{\underset{J=J_{1}}{\sum }}\left( \frac{2P_{J_{1}-\frac{1}{2}}\left(
x^{0}\right) }{(J_{1})(2J_{1}+1)}-\frac{2P_{J_{2}}\left( x^{0}\right) }{%
(J_{2}+1)(2J_{2}+1)}\right) =0  \label{nulflux}
\end{equation}%
where we have used $\int d^{3}xP_{J}\left( x^{0}\right) =4\pi \theta
^{3}(2J+2)\left( 2J+1\right) .$The formula (\ref{nulflux}) shows clearly
that the contributions to the divergence of the vector field\ $%
\overrightarrow{F}$ to the integral\ come from the boundaries only i.e. the
two fuzzy spheres of radii $J_{1}$ and $J_{2}$. This suggests that each of
the two\ terms in (\ref{nulflux}) represent the flux (of magnitude $-4\pi
\theta $ ) through the fuzzy sphere $J_{1,2}$ independently from the radii $%
J_{i}$. The opposite signs represent the directions of the fluxes; entering
or outgoing flux in the volume bounded by the two spheres of radii $J_{1,2}$%
. The fact that the result of the integral is zero shows the absence of a
charge between the spheres of radii $J_{1,2}\neq 0$ \ which express the
absence of singularities of $\overrightarrow{F}$ in this integration volume.
In what follows we will see that the singularity of the $U(1)$ instanton
field $\overrightarrow{F}$ at\ the origin is in fact due to a charge.

Each term in (\ref{nulflux}) can be represented by a flux of $F_{J}^{i}$
through a fuzzy sphere of radius $J$ as%
\begin{eqnarray}
-\frac{4\pi }{4\theta ^{2}}\int \left( x^{0}\right) ^{2}dx^{0}\frac{2P_{J-%
\frac{1}{2}}\left( x^{0}\right) }{J(2J+1)} &=&-\frac{4\pi }{4\theta ^{2}}%
\int x^{0}dx^{0}\frac{2x^{0}P_{J-\frac{1}{2}}\left( x^{0}\right) }{J(2J+1)} 
\notag \\
&=&\frac{4\pi }{4\theta }\int x^{0}dx^{0}-\frac{1}{4\theta }\frac{8\theta
^{2}J\left( J+1\right) P_{J}\left( x^{0}\right) }{J(J+1)(2J+1)}
\label{classicflux} \\
&=&\frac{1}{2\theta }\int \frac{d^{3}x}{x^{0}}\left( x^{i}\ast
F_{J}^{i}+F_{J}^{i}\ast x^{i}\right) =-4\pi \theta  \notag
\end{eqnarray}%
where we have used $x^{0}P_{J-\frac{1}{2}}\left( x^{0}\right) =2J\theta
P_{J}\left( x^{0}\right) ,$ $x^{i}\ast x^{i}\ast P_{J}\left( x^{0}\right)
=x^{0}\ast \left( x^{0}+2\theta \right) \ast P_{J}\left( x^{0}\right)
=4\theta ^{2}J\left( J+1\right) P_{J}\left( x^{0}\right) $ and $\int
x^{0}dx^{0}P_{J}\left( x^{0}\right) =\theta ^{2}\left( 2J+1\right) .$ The
third term of the right hand side of (\ref{classicflux}) represents the
analog of the classical flux of $F_{J}^{i}$ through the fuzzy sphere of
radius $J$. In what follows we will establish the noncommutative analog of
the Gauss theorem for any vector $\overrightarrow{V}(x)\in \mathcal{A}%
_{\theta }^{3}$ to justify the form of the third term of the right hand side
of (\ref{classicflux}). For this end we calculate the flux in the operator
formalism. In terms of operators, the integration (\ref{divintegral}) over a
volume bounded by a fuzzy sphere of radius $J$ is represented by a trace
over the Hilbert space representation $\mathcal{H}_{0}\oplus ...\oplus $\ $%
\mathcal{H}_{J}$ as 
\begin{eqnarray}
\Phi &=&\frac{\left( 2\pi \theta \right) ^{2}}{\pi }\overset{J}{\underset{j=0%
}{\sum }}\underset{m=-j}{\overset{j}{\sum }}\left\langle j,m\right\vert 
\frac{1}{2\theta }\left[ \widehat{\overline{z}}_{\alpha },\widehat{V}_{%
\overline{\alpha }}\right] -\frac{1}{2\theta }\left[ \widehat{z}_{\alpha },%
\widehat{V}_{\alpha }\right] \left\vert j,m\right\rangle \\
&=&2\pi \theta \underset{m=-J}{\overset{J}{\sum }}\left\langle
J,m\right\vert \widehat{\overline{z}}_{\alpha }\widehat{V}_{\overline{\alpha 
}}+\widehat{V}_{\alpha }\widehat{z}_{\alpha }\left\vert J,m\right\rangle
\label{tracediv}
\end{eqnarray}%
here the same mechanism of cancellation occurs for $j<J$ to leave only the
contribution of the boundary term which is a trace over $\mathcal{H}_{J}.$

$\widehat{\overline{z}}_{\alpha }\widehat{V}_{\overline{\alpha }}+\widehat{V}%
_{\alpha }\widehat{z}_{\alpha }$ can now be translated in terms of c-number
function as: 
\begin{equation}
\overline{z}_{\alpha }\ast V_{\overline{\alpha }}+V_{\alpha }\ast z_{\alpha
}=\overline{z}_{\alpha }V_{\overline{\alpha }}+\left( 2\theta \right) \frac{%
\partial }{\partial z_{\alpha }}V_{\overline{\alpha }}+V_{\alpha }z_{\alpha
}+\left( 2\theta \right) \frac{\partial }{\partial \overline{z}_{\alpha }}%
V_{\alpha }.  \label{DIVS}
\end{equation}%
By using (\ref{vector transform}) and (\ref{DIVV}) we can rewrite (\ref{DIVS}%
) in terms of star-product in $\mathcal{A}_{\theta }^{3}$ as%
\begin{equation*}
x^{i}V_{i}+V_{i}x^{i}+x^{0}\partial _{i}V_{i}=x^{i}\ast V_{i}+V_{i}\ast x^{i}
\end{equation*}%
which corresponds in terms of operators to $\widehat{x}^{i}\widehat{V}_{i}+%
\widehat{V}_{i}\widehat{x}^{i}\in $ $\widehat{\mathcal{A}}_{\theta }^{3}$.
Then the right hand side of (\ref{tracediv}) can be rewritten as 
\begin{eqnarray}
\Phi &=&2\pi \theta \underset{m=-J}{\overset{J}{\sum }}\left\langle
J,m\right\vert \widehat{x}^{i}\widehat{V}_{i}+\widehat{V}_{i}\widehat{x}%
^{i}\left\vert J,m\right\rangle =2\pi \theta Tr_{\mathcal{H}_{J}}\left( 
\widehat{x}^{i}\widehat{V}_{i}+\widehat{V}_{i}\widehat{x}^{i}\right) \\
&=&2\pi \theta Tr_{\mathcal{H}}\left( \widehat{x}_{J}^{i}\widehat{V}_{i}^{J}+%
\widehat{V}_{i}^{J}\widehat{x}_{J}^{i}\right)  \label{fluxv}
\end{eqnarray}%
which is the flux of the vector field $\widehat{V}$ through the fuzzy sphere
of radius $J$. \ $\widehat{x}_{J}^{i}=\widehat{P}_{J}\widehat{x}_{J}^{i}%
\widehat{P}_{J}$ and \ $\widehat{V}_{i}^{J}=\widehat{P}_{J}\widehat{V}%
_{i}^{J}\widehat{P}_{J}$ belong to the fuzzy sphere \ subalgebra $\widehat{S}%
_{\theta ,J}^{2}=\widehat{P}_{J}\widehat{\mathcal{A}}_{\theta }^{3}\widehat{P%
}_{J}\subset \widehat{\mathcal{A}}_{\theta }^{3}\subset \widehat{\mathcal{A}}%
_{\theta }^{4}$.

Note that we can use the correspondence (\ref{normalcorrespond}) to
translate (\ref{vector transform}) in terms of operators as%
\begin{equation*}
\widehat{V}_{\alpha }\left( \widehat{z},\widehat{\overline{z}}\right) =%
\widehat{V}_{i}\left( \widehat{x}\right) \frac{1}{2}\tau _{\alpha \beta }^{i}%
\widehat{\overline{z}}_{\beta },\text{ \ \ }\widehat{V}_{\alpha }\left( 
\widehat{z},\widehat{\overline{z}}\right) =\frac{1}{2}\widehat{z}_{\beta
}\tau _{\beta \alpha }^{i}\widehat{V}_{i}\left( \widehat{x}\right)
\end{equation*}%
Then from (\ref{tracediv}), we get directly (\ref{fluxv}) by using the
commutation relations of the coordinates $\widehat{z}_{\alpha }$ and $%
\widehat{\overline{z}}_{\alpha },$ $Tr\left( \tau \right) =0$ and (\ref%
{reduction}).

By combining (\ref{divintegral}) with (\ref{fluxv}) translated in term of
integral of $x_{J}^{i}\ast V_{i}^{J}+V_{i}^{J}\ast x_{J}^{i}$ over a fuzzy
sphere of radius $J$, we can establish the noncommutative analog of Gauss
theorem for any vector field $\overrightarrow{V}=\overset{\infty }{\underset{%
J=0}{\sum }}P_{J}\ast \overrightarrow{V}\ast P_{J}=\overset{\infty }{%
\underset{J=0}{\sum }}\overrightarrow{V}^{J}\in \mathcal{A}_{\theta }^{3}$ as

\begin{equation*}
\int d^{3}x\underset{J=0}{\overset{J_{1}}{\sum }}\partial _{i}V_{i}^{J}=%
\frac{1}{2\theta }\int \frac{d^{3}x}{x^{0}}(x_{J_{1}}^{i}\ast
V_{i}^{J_{1}}+V_{i}^{J_{1}}\ast x_{J_{1}}^{i})
\end{equation*}

where the left hand side is the integral of the divergence of $%
\overrightarrow{V}$ over a volume in $\mathbb{R}_{\theta }^{3}$ bounded by a
fuzzy sphere of radius $J_{1}$ and the right hand side is a surface term
which is the integral over the fuzzy sphere of radius $J_{1}$ which express
the flux of the vector field $\overrightarrow{V}\in \mathcal{A}_{\theta
}^{3} $ through this sphere.

Applying the above formalism to the case of our strength field $\widehat{F}%
^{i}$, we get from (\ref{fluxv}) the flux through the fuzzy sphere as 
\begin{eqnarray*}
\Phi &=&2\pi \theta Tr_{\mathcal{H}_{J}}\left( \widehat{x}^{i}\widehat{F}%
^{i}+\widehat{F}^{i}\widehat{x}^{i}\right) =2\pi \theta Tr_{\mathcal{H}%
}\left( \frac{-2\widehat{x}_{i}\widehat{x}_{i}\widehat{P}_{J}}{4\theta
^{2}J\left( J+1\right) \left( 2J+1\right) }\right) \\
&=&4\pi \theta Tr_{\mathcal{H}_{J}}\left( \frac{-J\left( J+1\right) }{%
J\left( J+1\right) \left( 2J+1\right) }\right) =-4\pi \theta \underset{m=-J}{%
\overset{J}{\sum }}\frac{1}{(2J+1)}=-4\pi \theta .
\end{eqnarray*}

This result shows that the Flux $\Phi $ is independent from the choice of
the representation $J$ i.e. it is independent from the choice of the sphere
radius 
\begin{equation}
\Phi =-4\pi \theta \text{ \ }\forall J\text{ the radius of the fuzzy sphere.}
\label{Gauss}
\end{equation}%
This is similar to the classical \emph{Gauss theorem} which states that 
\emph{the total charge calculated by the flux of a coulombian or Dirac
monopole field through a closed surface is independent from the shape of
this closed surface.}

Thus (\ref{Gauss}) is the noncommutative analog of the Gauss theorem on the
noncommutative space $\mathbb{R}_{\theta }^{3}$

This result shows also that the quantity $\Phi $ represents a source of the
field $\widehat{\overrightarrow{F}}$ which we can be interpreted as a
magnetic charge, Thus the $U(1)$-one-instanton on $\mathbb{R}_{\theta }^{4}$
gives rise to a magnetic charge at the origin of the noncommutative space $%
\mathbb{R}_{\theta }^{3}$.

\subsection{$\mathcal{H}_{KL}$-Instanton solution Flux through the fuzzy
sphere and gauge invariance}

In this section we try to see in what sense the magnetic\ flux is invariant
under\ partial\ isometries. First we rewrite the transformation $U$ in the
Schwinger basis as 
\begin{equation}
U=\overset{\infty }{\underset{J=0}{\sum }}\underset{m=-J}{\overset{J}{\sum }}%
\left\vert J,m\right\rangle \left\langle J+P,m+Q\right\vert =\overset{\infty 
}{\underset{J=P}{\sum }}\overset{J-P+Q}{\underset{m=-\left( J-P\right) +Q}{%
\sum }}\left\vert J-P,m-Q\right\rangle \left\langle J,m\right\vert
\label{iso}
\end{equation}%
where $P=\frac{1}{2}\left( K+L\right) ,$ $Q=\frac{1}{2}\left( K-L\right) $.
Then the components of the transformed field is given by

\begin{equation}
\widehat{\overrightarrow{F}}_{KL}=U^{\dagger }\widehat{\overrightarrow{F}}%
_{00}U=\widehat{\overrightarrow{F}}_{u}=-\frac{\theta }{\widehat{x}_{u}^{0}(%
\widehat{x}_{u}^{0}+2\theta )(2\widehat{x}_{u}^{0}+2\theta )}\widehat{%
\overrightarrow{x}}_{u}  \label{F3U}
\end{equation}

where

\begin{equation}
\widehat{x}_{u}^{i}=U^{\dagger }\widehat{x}^{i}U.\text{and }\widehat{x}%
_{u}^{0}=U^{\dagger }\widehat{x}^{0}U.  \label{ISOX}
\end{equation}

These new coordinates $\widehat{x}_{u}^{i}$ satisfy the same commutation
relations as the old ones%
\begin{equation*}
\left[ \widehat{x}_{u}^{i},\widehat{x}_{u}^{j}\right] =i2\theta \epsilon
^{ijk}\widehat{x}_{u}^{k},\text{ \ }\left[ \widehat{x}_{u}^{0},\widehat{x}%
_{u}^{i}\right] =0.
\end{equation*}%
and

\begin{equation*}
\widehat{x}_{u}^{i}\widehat{x}_{u}^{i}=\widehat{x}_{u}^{0}(\widehat{x}%
_{u}^{0}+2\theta )
\end{equation*}%
Note that in this case the formalism is expressed in the subalgebra $%
\widehat{\mathcal{A}}_{\theta ,u}^{3}=U^{\dagger }\widehat{\mathcal{A}}%
_{\theta }^{3}U\subset \widehat{\mathcal{A}}_{\theta }^{3}\subset \widehat{%
\mathcal{A}}_{\theta }^{4}$. The formula (\ref{ISOX}) is similar to the
transformations of the angular momentum in quantum mechanics\footnote{%
The difference between the quantum mechanical\ gauge transformation\ and the
isometry we discuss in our present paper is that\ unitary gauge
transformations in quantum mechanics just rotate the states $\left\vert
Jm\right\rangle $, whereas in our case the isometry (which is a gauge
transformation) generate a shift of $J$ then a rotation of the same basis
vectors $\left\vert Jm\right\rangle $.}. These transformations are
concretely given by 
\begin{eqnarray*}
\widehat{x}_{u}^{3} &=&\overset{\infty }{\underset{J=P}{\sum }}\overset{J-P+Q%
}{\underset{m=-\left( J-P\right) +Q}{\sum }}2\theta \left( m-Q\right)
\left\vert J,m\right\rangle \left\langle J,m\right\vert , \\
\widehat{x}_{u}^{+} &=&\overset{\infty }{\underset{J=P}{\sum }}\overset{J-P+Q%
}{\underset{m=-\left( J-P\right) +Q}{\sum }}2\theta \sqrt{\left( J-P\right)
\left( J-P+1\right) -\left( m-Q\right) \left( m-Q+1\right) }\left\vert
J,m+1\right\rangle \left\langle J,m\right\vert , \\
\widehat{x}_{u}^{-} &=&\overset{\infty }{\underset{J=P}{\sum }}\overset{J-P+Q%
}{\underset{m=-\left( J-P\right) +Q}{\sum }}2\theta \sqrt{\left( J-P\right)
\left( J-P+1\right) -\left( m-Q\right) \left( m-Q-1\right) }\left\vert
J,m-1\right\rangle \left\langle J,m\right\vert , \\
\widehat{x}_{u}^{0} &=&\overset{\infty }{\underset{J=P}{\sum }}\overset{J-P+Q%
}{\underset{m=-\left( J-P\right) +Q}{\sum }}2\theta \left( J-P\right)
\left\vert J,m\right\rangle \left\langle J,m\right\vert .
\end{eqnarray*}

As in the above section, we can use the projector on the fuzzy sphere $%
\widehat{P}_{J}$ ,with $J=J^{^{\prime }}+P$, to define the map (surjection): 
$H_{J}\longrightarrow H_{J^{^{\prime }}}$

\bigskip 
\begin{equation*}
U_{J}=U\widehat{P}_{J}=\overset{J-P+Q}{\underset{m=-\left( J-P\right) +Q}{%
\sum }}\left\vert J-P,m-Q\right\rangle \left\langle J,m\right\vert
\end{equation*}

satisfying\newline
\begin{equation*}
U_{J}^{\dagger }U_{J}=\overset{J-P+Q}{\underset{m=-\left( J-P\right) +Q}{%
\sum }}\left\vert J,m\right\rangle \left\langle J,m\right\vert
\end{equation*}

which is a projector on a part of states of the fuzzy sphere of radius $J$
and

\begin{equation*}
U_{J}U_{J}^{\dagger }=\overset{j-P+Q}{\underset{m=-\left( j-P\right) +Q}{%
\sum }}\left\vert j-P,m-Q\right\rangle \left\langle j-P,m-Q\right\vert
.=P_{J^{^{\prime }}}
\end{equation*}%
is a projector on the fuzzy sphere of radius $J^{^{\prime }}$. Under $U_{J}$%
, The coordinates $\widehat{x}^{i}$ transform as

\begin{equation*}
\widehat{x}_{u,J}^{i}=U_{J}^{\dagger }\widehat{x}^{i}U_{J}=\widehat{P}_{J}%
\widehat{x}_{u}^{i}\widehat{P}_{J}.
\end{equation*}

They satisfy the commutation rules on the fuzzy sphere%
\begin{equation*}
\left[ \widehat{x}_{u,J}^{i},\widehat{x}_{u,J}^{j}\right] =i2\theta \epsilon
^{ijk}\widehat{x}_{u,J}^{k},\text{ \ }\left[ \widehat{x}_{u,J}^{0},\widehat{x%
}_{u,J}^{i}\right] =0.
\end{equation*}%
and

\begin{equation*}
\widehat{x}_{u,J}^{i}\widehat{x}_{u,J}^{i}=\widehat{x}_{u,J}^{0}(\widehat{x}%
_{u,J}^{0}+2\theta ),
\end{equation*}

where $\widehat{x}_{u,J}^{0}=U_{J}^{\dagger }\widehat{x}^{0}U_{J}$.

The transformed coordinates generate the subalgebra $\widehat{S}_{u,J}^{2}$
which is an algebra describing a truncated fuzzy sphere realized in the
Hilbert subspace $\mathcal{H}_{u,\unit{J}}\subset \mathcal{H}_{J}$ spanned
by the basis $\left\vert J,m-Q\right\rangle $ with $J=J^{^{\prime }}+P$ and $%
-$ $J^{^{\prime }}\leq m\leq J^{^{\prime }}$. In this truncated space $%
\mathcal{H}_{u,\unit{J}}$ the coordinates $\widehat{x}_{u,J}^{i}$ act as

\begin{eqnarray*}
\widehat{x}_{u,J}^{3}\left\vert J,m\right\rangle &=&2\theta \left(
m-Q\right) \left\vert J,m\right\rangle , \\
\widehat{x}_{u,J}^{+}\left\vert J,m\right\rangle &=&2\theta \sqrt{\left(
J-P\right) \left( J-P+1\right) -\left( m-Q\right) \left( m-Q+1\right) }%
\left\vert J,m+1\right\rangle , \\
\widehat{x}_{u,J}^{-}\left\vert J,m\right\rangle &=&2\theta \sqrt{\left(
J-P\right) \left( J-P+1\right) -\left( m-Q\right) \left( m-Q-1\right) }%
\left\vert J,m-1\right\rangle
\end{eqnarray*}

which show that the highest state is $\left\vert J,J-P+Q\right\rangle $ ($%
\widehat{x}_{Ju}^{+}\left\vert J,J-P+Q\right\rangle =0)$ and the lower state
is $\left\vert J,-(J-P)+Q\right\rangle $ ($\widehat{x}_{Ju}^{-}\left\vert
j,-(j-P)+Q\right\rangle =0)$.

Now this formalism is expressed in the subalgebra $\widehat{S}%
_{u,J}^{2}\subset \widehat{\mathcal{A}}_{\theta ,u}^{3}\subset \widehat{%
\mathcal{A}}_{\theta }^{3}\subset \widehat{\mathcal{A}}_{\theta }^{4}.$ The
calculus of the flux is now given by a trace over $\mathcal{H}_{u,\unit{J}}$.%
\begin{eqnarray*}
\Phi &=&2\pi \theta Tr_{\mathcal{H}_{u,J}}\left( \widehat{x}_{u,J}^{i}%
\widehat{F}_{u,J}^{i}+\widehat{F}_{u,J}^{i}\widehat{x}_{u,J}^{i}\right)
=-2\pi \theta Tr_{\mathcal{H}_{u,J}}\frac{2\theta \widehat{x}_{u,J}^{i}%
\widehat{x}_{u,J}^{i}}{\widehat{x}_{u}^{0}(\widehat{x}_{u}^{0}+2\theta )(2%
\widehat{x}_{u}^{0}+2\theta )} \\
&=&\frac{-2\left( 2\pi \theta \right) }{\left( 2J-2P+1\right) }\underset{%
m=-\left( J-P\right) +Q}{\overset{J-P+Q}{\sum }}1=-4\pi \theta
\end{eqnarray*}

Thus we obtain the same result as in (\ref{Gauss}) and also in this gauge
the flux of the transformed strength field is independent from the choice of
the radius i.e. independent from the Hilbert space representation $\mathcal{H%
}_{u,\unit{J}}$%
\begin{eqnarray*}
\Phi &=&-4\pi \theta \text{ \ }\forall J>P\text{ the radius of the fuzzy
sphere, }\forall \text{ }K,L\geq 0 \\
\text{or }P &=&0,\frac{1}{2},1,...\infty \,\text{\ and }Q\ \text{runs by
integer step over }-P\leq Q\leq P.
\end{eqnarray*}%
\ 

\begin{quotation}
\textbf{Discussion and conclusion}
\end{quotation}

In this paper we have investigated $U(1)$-one-instanton solutions with the
intension of studying the relation between the instanton number and partial
isometries. This allows us to see that the partial isometry transformations
which act like $U(1)$ gauge transformations leave invariant the instanton
number

The second result, obtained in this paper, is that the description of the $%
U(1)$-one-instanton solution in terms of $\widehat{\mathcal{A}}_{\theta
}^{3} $ algebra, gives rise to an object behaving like Dirac magnetic
monopole field. Its flux through the fuzzy spheres\ is independent from the
choice of the radius of these ones. This fact can be seen as the
noncommutative analog of the Gauss theorem for the coulombian forces in the
classical case. Furthermore we have seen that the flux is invariant under
the partial isometry transformations of the $U(1)$ instanton field, provided
that the coordinates on $\mathbb{R}_{\theta }^{3}$ transform as $\widehat{x}%
_{u}^{i}=U^{\dagger }\widehat{x}^{i}U$. This transformation preserve the
commutation rules of the coordinates algebra\ on $\mathbb{R}_{\theta }^{3}$,
and confirm the mixture between gauge theory and geometrical transformations
acting\ on the base space, which, in our case, rotate the fuzzy sphere and
shifts the radius.

Finally we hope to comment the fact that the flux $\Phi $ can be related to
the quantization of the magnetic charge; This can be done by remarking that
the action of the $\mathbb{R}_{\theta }^{3}$ coordinates is the same as the
quantum mechanical angular momentum. Thus we have in one part :%
\begin{gather*}
J^{3}\left\vert J,m\right\rangle =m\hbar \left\vert J,m\right\rangle \\
J^{\pm }\left\vert J,m\right\rangle =\hbar \sqrt{J\left( J+1\right) -m\left(
m\pm 1\right) }\left\vert J,m\pm 1\right\rangle
\end{gather*}%
and\ in the other part :%
\begin{gather*}
\widehat{x}^{3}\left\vert J,m\right\rangle =m\left( 2\theta \right)
\left\vert J,m\right\rangle \\
\widehat{x}^{\pm }\left\vert J,m\right\rangle =\left( 2\theta \right) \sqrt{%
J\left( J+1\right) -m\left( m\pm 1\right) }\left\vert J,m\pm 1\right\rangle
\end{gather*}%
So we can identify $\hbar $ with $\left( 2\theta \right) $, then the flux
will be given by $\Phi =-2\pi \hbar $, which correspond to a monopole\
magnetic\ charge $-1$.

As further directions one can generalize the statements above on
noncommutative\ multi-$U(1)$-instanton solutions on $\mathbb{R}_{\theta
}^{4} $ by \ searching for\ general solutions to calculate their flux
through fuzzy spheres to hope to find higher monopole charges.

\textbf{Acknowledgment}

I would like to thank M. Dubois-Violette for hepful discussion.

\end{document}